\documentclass{aa}
\usepackage{graphicx}
\usepackage[varg]{txfonts}
\usepackage{longtable}

\usepackage{natbib}

\def\ltsima{$\; \buildrel < \over \sim \;$}
\def\simlt{\lower.5ex\hbox{\ltsima}}
\def\gtsima{$\; \buildrel > \over \sim \;$}
\def\simgt{\lower.5ex\hbox{\gtsima}}
\def\AA{\; \buildrel \circ \over {\rm A}}

\begin{document}
   \title{An updated survey of globular clusters in M31. II.}
\subtitle{Newly discovered bright and remote clusters.}

   \author{S. Galleti\inst{1,2}, M. Bellazzini\inst{2}, L. Federici\inst{2},
           A. Buzzoni\inst{2}
          \and
       F. Fusi Pecci\inst{2}
\thanks{Based on observations made with the Italian Telescopio Nazionale Galileo
 (TNG) operated on the island of La Palma by the Fundaci\'on Galileo Galilei of
 the INAF (Istituto Nazionale di Astrofisica) at the Spanish Observatorio del
 Roque de los Muchachos of the Instituto de Astrofisica de Canarias.
 Based on observations made with the Cassini Telescope (Loiano,
 Italy). The Cassini telescope is operated by INAF - Osservatorio Astronomico di
 Bologna.}
          }

   \offprints{S. Galleti}

   \institute{Universit\`a di Bologna, Dipartimento di Astronomia
             Via Ranzani 1, 40127 Bologna, Italy\\
            \email{silvia.galleti2@unibo.it}
         \and
             INAF - Osservatorio Astronomico di Bologna,
              Via Ranzani 1, 40127 Bologna, Italy\\
          \email{michele.bellazzini@oabo.inaf.it, luciana.federici@oabo.inaf.it, \\
          alberto.buzzoni@oabo.inaf.it, flavio.fusipecci@oabo.inaf.it}             }

     \authorrunning{S. Galleti et al.}
   \titlerunning{A spectroscopic survey of globular clusters in M31. II. Remote
   clusters.}

   \date{Submitted 4 May 2007; Accepted 22 May 2007 }

\abstract
{}
{We present the first results of a large spectroscopic survey of candidate
globular clusters located in the extreme outskirts  of the nearby M31 galaxy.
The survey is aimed at ascertaining the nature of the selected candidates to
increase the sample of confirmed M31 clusters lying more that $2\degr$ away
from the center of the galaxy.}
{We obtained low resolution spectra
($\lambda/\Delta\lambda \simeq 800 - 1300$) of 48 targets selected from the
Extended Source Catalogue of 2MASS, as in Galleti et al. (2005).  The observed
candidates have been robustly classified according to their radial velocity
and by verifying their extended/point-source nature from ground-based optical
images. We have also obtained a spectrum and a
radial velocity estimate for  the remote M31 globular discovered
by Martin et al. (2006b).}
{Among the 48 observed candidates clusters we found: 35
background galaxies, 8 foreground Galactic stars, and 5 genuine remote globular
clusters. One of them has been already identified  independently by Mackey et
al. (2007), their GC1; the other four are completely new discoveries: B516,
B517, B518, B519. The newly discovered clusters lie at projected distance  40
kpc $\simlt R_p\simlt$ 100 kpc from the center of M31, and have absolute
integrated magnitude $ - 9.5 \simlt M_V\simlt -7.5$. For all the observed
clusters we have measured the
strongest Lick indices and we have obtained spectroscopic metallicity
estimates. Mackey-GC1, Martin-GC1, B517 and B518 have spectra typical of old
and metal poor globular clusters ([Fe/H]$\simlt -1.3$); B519 appears old but
quite metal-rich ([Fe/H]$\simeq -0.5$); B516 presents very strong Balmer
absorption lines: if this is indeed a cluster it should have a relatively
young age  (likely $<2$ Gyr).
}
{The present analysis nearly doubles the number of M31 globulars at
$R_p\ge$ 40 kpc. At odds with the Milky Way, M31 appears to have a
significant population of very bright globular clusters in its
extreme outskirts.
}

   \keywords{Galaxies: individual: M~31 -- Galaxies: star clusters --
    catalogs --- Galaxies: Local Group          }

   \maketitle
%

\section{Introduction}

The study of globular clusters (GC) systems is a fundamental astrophysical
tool to
investigate the formation and the evolution of distant galaxies. GCs are
ubiquitous and abundant in virtually any type of galaxy;  they are
intrinsically bright and can be identified even at large distances; their
integrated colors and spectra are relatively easy to interpret, since they
typically host a Simple Stellar Population (Renzini \& Fusi Pecci \cite{rfp88});
their
kinematics are powerful probes for the gravitational potential of their parent
galaxy (see Brodie \& Strader \cite{brodie}, and references therein).

In this context, the GC system of our next neighbor spiral galaxy,
M31, plays a key and twofold role:

\begin{itemize}

\item It is the richest GC system that we can study with the same
integrated-light methods that we apply to any distant galaxy, and whose
individual GCs can also be resolved into stars with HST. This
provides a fundamental sanity check for our observations
of distant GC systems.

\item It has a much larger number of members ($\sim 475 \pm 25$,
as estimated by Barmby et al. \cite{bar01} with respect to
the GC system  of the Milky Way ($\simeq 150$ GCs),
and, apparently, a larger variety of cluster ``species'' (including, for
example, the massive young disc clusters (BLCC) described by
Fusi Pecci et al. \cite{ffp}, or the Extended Clusters (EC) discovered by
Huxor et al. \cite{huxext}).
This provides the opportunity to study in detail
systems that are rare or have no counterpart in our own Galaxy.

\end{itemize}

In spite of uninterrupted study since the times of Hubble \cite{hubble},
we are still lacking a complete knowledge of the GC system of M31
(Barmby et al. \cite{bar01}, Galleti
et al. \cite{G06a}, hereafter G06a).
At present, we know more than 350 confirmed members,
but hundreds of candidates are still to be checked (G06a) and any new
survey finds out new clusters or promising candidates
(see, for example, Mochejska et al. \cite{moche}; Huxor et al. \cite{hux},
hereafter H04; G06a and references therein).
In particular, some recent studies (H04; Galleti et al. \cite{b514},
hereafter G05; Martin et al. \cite{martin}, hereafter M06b)
have opened a window on a realm that was completely unexplored:
that of remote M31 clusters, i.e. those lying at a projected distance
($R_p$) larger than 30 - 40 kpc from the center of the galaxy.

Until two years ago, the outermost M31 cluster known was G1, located at
$R_p\simeq 35$ kpc. This situation was quite disappointing,
since in the Milky Way there are seven clusters lying at $R_{GC}>40$ kpc, while
M31 - which, as said, has a much richer GC system - seemed to have none.
In G05 we presented a method to select candidate remote GCs in M31 from the
Extended Sources Catalogue of 2MASS (Skrutskie et al. \cite{skrut}).
The nature of the selected candidates must be subsequently ascertained by means
of low resolution spectra that provide the radial velocity estimate by which a
genuine globular cluster can be told from background galaxies or foreground
stars (see G06a and references therein). In G05 we also presented the
spectroscopic follow up of two of our candidates that lead to the discovery of
the outermost cluster of M31, B514, located at $R_p\simeq 55$ kpc from the
center of M31. The observed spectra indicated old age and low metallicity for
B514. Subsequent follow up with HST ACS/WFC allowed us to obtain
a deep Color Magnitude Diagram (CMD), confirming B514 as a genuine old and
metal-poor, bright GC ($M_V\simeq -9.1$), with a blue Horizontal Branch
(Galleti et al. \cite{G06b}, hereafter G06b).
Later, M06b identified an even more extreme
cluster at $R_p\sim 118$ kpc; very recently, Mackey et al. \cite{mackhux1}
presented ACS/WFC photometry of ten
additional clusters
from H04, three of them lying at $R_p> 40$ kpc\footnote{Plus B514 that they
re-observed, and that is named GC4 in their list.}.

Here we present the results of the first year of our ongoing survey
for remote M31 globular clusters whose earliest results were
described in G05\footnote{In parallel we are carrying on a survey to
ascertain the nature of already known CGCs, drawn from the Revised
Bologna Catalogue (G06a), see {\tt http://www.bo.astro.it/M31/}.}; 
we report on the discovery of four new
bright globular clusters located at projected distance 
40~kpc~$\simlt R_p\simlt$~100~kpc from the center of M31.

All over the paper we will adopt D=783 kpc for M31, from McConnachie et al.
\cite{mcc}; at this distance, $1\degr$ corresponds to $\simeq 13.7$ kpc,
$1\arcmin$ to $\simeq 228$ pc. $V_r=-301$ km/s is adopted as the systemic radial
velocity of M31, as in G06a.

\subsection{Selection of candidate remote clusters}

The catalogues from the 2MASS all-sky survey (Skrutskie et al.
\cite{skrut}) provide the possibility of selecting interesting
sources in arbitrarily large areas of the sky around M31. In Galleti
et al. \cite{silvia} - hereafter G04 - we searched for the
counterparts of known M31 GCs and candidates GCs (CGCs) in 2MASS and
we found that the large majority (92\%) of genuine GCs are included
in the Point Source Catalogue (PSC), while only the remaining 8\% is
in the Extended Source Catalogue (XSC). This statistic suggests that
if a candidate M31 GC is included in the XSC it is very likely a
genuine extended source; in principle, the selection of candidates
from the XSC should overcome any contamination  from foreground
stars. Therefore, a CGC selected from the XSC can {\em only} be a
genuine GC {\em or} a distant galaxy in the background: a simple low
resolution spectrum is sufficient to tell one case from the other
since M31 GCs have radial velocities around $V_r=-301 \pm 480$ km/s,
while background galaxies have cosmological recession velocities
($V_r>+5000$ km/s; G06a).

Actual selection criteria have evolved while the survey was ongoing and will
presumably be further refined and adjusted in the future.
Here we describe the preliminary selection procedure that has been
adopted until now.

First of all, we considered sources extracted from 2MASS-XSC located
within a $\sim 20\degr \times 20\degr$ area around M31 excluding the
innermost $2\degr$. We limited the extraction to well-behaved
sources that have valid measures of the magnitude in J, H, and K
(13006 sources). To reduce the number of eligible sources to a more
manageable number we impose three strong selection criteria,
tailored on the observed characteristic of confirmed M31 GCs:

\begin{enumerate}

\item GCs are quite round in shape, in general. M31 appear to host clusters of
larger ellipticity with respect to the Milky Way, still there is no known GC
having $e=1.-b/a>0.4$ (Barmby et al. \cite{bar02}).
Since the XSC provides the axis ratio ($b/a$) for all
the listed sources we can impose a limit in  $e$ to the selected sources. To
exclude from the sample all the disk galaxies seen with a significant
inclination and very elongated elliptical galaxies we retained only sources
having $e\le 0.4$.

\item A very broad limit in non reddening-corrected color ($J-K\le 1.2$)
was adopted to exclude the
reddest early type galaxies, that are the most abundant contaminants
in the sample.

\item The characteristic size (half-light-radius $r_h$) of globular clusters
at the distance of M31 is $0.3\arcsec \sol r_h\sol 9\arcsec$
(Barmby et al. \cite{bar02}).
To reject relatively nearby background galaxies that may have large projected
sizes we excluded all the sources having
characteristic radius measured in the J images, $R_J>10\arcsec$. We chose $R_J$
because J images have the highest S/N in 2MASS, therefore the most reliable
measure of the size of an object is obtained in this passband.

\end{enumerate}

Using he above criteria, we selected $\sim$1800 objects that were finally
submitted
to direct visual inspection on DSS2 images\footnote{
The DSS2 images were obtained from the site
{\tt http://archive.eso.org/dss/dss}.}.
In this phase we mainly rejected
nearly-face-on spirals and/or irregular galaxies that were not pruned by
automatic criteria. The visual analysis of possible candidates is currently
ongoing. At present we have inspected the images of the first 280 of them
(randomly chosen):
109 passed the ``visual inspection test'' and were
retained in the final list of {\em good candidates} that
deserve spectroscopic follow-up. Here we describe the follow-up of 48
candidates. If we include the results of the pilot project (G05), at the present
stage of advancement of the survey we have obtained spectra of 50 candidates
and we obtained a harvest of six bona-fide remote M31 GCs.
Two of them were
independently discovered also by H04 (see also Mackey et al. \cite{mackhux1}),
four are completely new discoveries. The global success rate of our selection
procedure is 12\%.

\subsubsection{Selection biases}

From the above description of our selection criteria it is quite
clear that our final sample is subject to several biases. By
definition we cannot find very elliptical, very red and/or very
extended clusters (as those identified by Huxor et al. \cite{huxext}).
As we draw our
candidates from the XSC, we cannot select any cluster appearing as
point-like in 2MASS (G04). A limit in integrated luminosity is
automatically imposed by the limiting magnitude of 2MASS\footnote{
The limiting magnitude of 2MASS data roughly corresponds to $M_V\sim -6.5$
in the Luminosity Function of M31 GCs.
Note also that we always used standard 2MASS data, not the recently released
``6x'' data
(http://www.ipac.caltech.edu/2mass/releases/allsky/doc/seca3\_1.html) that are
$\sim$ 1 mag deeper, but are limited to a central field of area 2.8 deg$^2$.};
moreover
we tend to favor brighter candidates when we perform the follow-up
observations, for obvious reasons of convenience. We clearly caution
the reader that the goal of the present survey is to find out as
many remote M31 GCs as possible, but the survey cannot be by no
means {\em complete} or fully {\em objective}. In spite of this, the
results demonstrate that it is quite useful in finding out members
of a population of M31 clusters that has remained completely hidden
until a couple of years ago.

H04 and M06b identified their remote M31 GCs from wide
and deep optical surveys (INT-WFC and MEGACAM surveys, see Ibata et al.
\cite{iba01,iba04} and M06b). They can certainly reach
fainter clusters and in many cases GCs are partially resolved in their images
(H04), hence quite easy to identify even without spectroscopic follow-up
(in fact, they are taking into consideration {\em only objects that are - at
least partially -  resolved into stars}).
On the other hand, 2MASS allows us to survey a much wider area of the sky,
including large
portions of the outskirts of M31 that are not reached by the INT-WFC and
MEGACAM surveys. In this sense it is interesting to note that three
of the four newly discovered clusters presented here lie in regions of the sky
not covered by these optical surveys, while one is outside of the INT-WFC area
but is included into the MEGACAM survey (see Sect.~3., below).

\section{Observations and data reduction}

A database of spectra for 48 targets has been assembled through several
observing runs, using different instruments (see below).
The main aim of
the survey is to obtain low resolution spectra of candidate M31 clusters
to measure their radial velocity (RV; $V_r$), that allows us to ascertain
if the considered candidate is a genuine GCs or a background galaxy, according
to the criteria discussed in G06a. In some cases we had also to verify the
extended nature of the candidates with supplementary imaging, using the
technique described in G06a (see below).

\subsection{DOLORES Data}

The imager/spectrograph DoLoRes at the 3.52m TNG telescope in La Palma
(Canary Island, Spain) was used in service mode in various nights
during the period
October - December, 2005 (Run 1), and in visitor mode in the nights of
October 10-15, 2006 (Run 2),
to acquire long slit spectra of 38 M31 CGCs.
DoLoRes is equipped with a 2048$\times$2048 px$^2$ thinned and back-illuminated Loral CCD
array with a total field of view of $9.4\arcmin \times 9.4\arcmin$.
The adopted MG-B grism yields a resolution of
$ 6 \AA$ (R=875) with $1\arcsec$ slit, and covers the spectral range
$3800 \AA < \lambda < 6800 \AA$. Our typical exposure times range from
15 to 20 minutes for Run 1 and from 45 to 60 minutes for Run 2,
 giving spectra with typical
$S/N \sim 35$ per resolution element. In both runs, high S/N spectra of the
bright clusters B158 and B225 were also obtained,
to be used as templates for the
estimates of $V_r$ as in G05 and G06a.

During Run 1, technical problems prevented the acquisition of a
reference lamp after each spectrum. For this reason the wavelength
calibration of science spectra was of low quality and the typical
uncertainty in radial velocity was $\sim \pm 200$ km/s. This allowed
a good discrimination between GCs and background galaxies but the
$V_r$ estimates of GCs were completely unusable for any other
purpose. The GCs identified during this run have been re-observed
later at the Cassini telescope (see below).

A He lamp spectrum was acquired after each science frame for
wavelength calibration during Run 2. Spectra of the adopted radial velocity
template clusters have been obtained during each observing night, to ensure the
self consistency of the $V_r$ scale over the whole run.
Bias subtraction, flat-fielding, and sky subtraction were performed using
standard packages in IRAF, as described in G05.

\subsection{BFOSC Data}

Long slit spectra for 16 CGCs in M31 were obtained with the low
resolution spectrograph BFOSC (Gualandi \& Merighi \cite{manbfosc})
mounted at the 1.52m Cassini Telescope of the Loiano Observatory,
near Bologna (Italy), during several runs in 2006: August 19-22 (Run
1), September 1-2 (Run 2), October 25-27 (Run 3), and November 22-23
(Run 4). In six cases we re-observed interesting candidates
identified during TNG-Run 1; in two cases we observed a given target
in two different runs to obtain a spectrum of sufficient S/N for
subsequent analysis. During Run 1 we acquired a spectrum of the
remote cluster recently discovered by M06b; in the following we
will refer to this clusters as to Martin-GC1.

Nearly 60\% of the nights were suitable
for useful observations.
The typical seeing was $1.5\arcsec-2.5\arcsec$ FWHM.
The detector was a thinned, back illuminated EEV CCD,
with $1300 \times 1340 ~px^2$.  A
 $1.5\arcsec$ slit was used. The adopted set-up  provides a spectral
resolution $\Delta \lambda = 4.1 \AA$ (${\lambda}\over{\Delta
\lambda} $ $\sim 1300$) and covers the range $4200 \AA < \lambda <
6600 \AA$. We took a He-Ar calibration lamp spectrum after each
scientific exposure, maintaining the same pointing of the telescope.
Integrations were typically 60 minutes per exposure, yielding
spectra with characteristic signal-to-noise ratio $S/N \sim 24$ per
resolution element. During each observing night we also observed,
with the same set-up, at least a couple of template targets that we
adopted as radial velocity standards. Such targets are extracted
from the list of bright M31 clusters (Master RV templates) for which
we have constructed very high S/N template spectra by stacking
several spectra obtained in several BFOSC observing runs (see
Appendix A for details of the procedure published only in the
electronic edition of the Journal). In practice the spectra of RV
standard that are obtained during each observing night are used to
ensure that the measures obtained during the night are in the RV
scale defined by the Master RV templates. The data reduction steps
were performed using standard packages in IRAF, as for the Dolores
spectra, above.

\begin{table*}
\begin{center}
\caption{Observed candidates globular cluster \label{oss}}
{
\begin{footnotesize}
\begin{tabular}{l c r r r r r r c c c}
\hline \hline
2Mass-Name& K & $ V_r$  &$ \pm\epsilon_{Vr}$& Source& $V_r$&$ \pm\epsilon_{Vr}$& Source& R &E/PS& Name\\
\hline
2MASX-J00255466+4057060 &  12.37&12200(EL$^a$)& 500& R1 TNG  &      &    &       & & &      \\
2MASX-J00264769+3944463 &  13.64&   -350$^1$  & 200& R1 TNG  &-219  & 15 & R1 LOI &1.27 &E&Mackey-GC1    \\
2MASX-J00285169+3905167 &  13.82&   -100$^1$  & 200& R1 TNG  &  22  & 14 & R1 LOI &1.01 &PS&       \\
2MASX-J00294581+3729300 &  11.97&   10800$^1$ &1200& R1 TNG  &11421 & 12 & R2 LOI & & &     \\
2MASX-J00301921+4117203 &  14.25& M Star($^b$)& 200& R1 TNG  &      &    &   & & &      \\
2MASX-J00305689+4520599 &  13.82&   19400     &1500& R1 TNG  &      &    &       & & &      \\
2MASX-J00313051+4033153 &  14.15& M Star      &38  & R2 TNG  &      &    &       & & &    \\
2MASX-J00313771+4034163 &  14.00&    52861    &121 & R2 TNG  &      &    &       & & &   \\
2MASX-J00350474+4235316 &  13.61&  19700$^1$  &1200& R1 TNG  &19638 & 43 & R2 LOI & & &    \\
2MASX-J00442168+3843319 &  14.34&   -150$^1$  &200 & R1 TNG  &-23   & 14 & R1 LOI &0.99 &PS&      \\
2MASX-J00491125+3806514 &  12.58&   11400(EL) &500 & R1 TNG  &      &    &       & & &    \\
2MASX-J00492851+3826166 &  14.22& M Star      & 200& R1 TNG  &      &    &       & & &    \\
2MASX-J00553861+4524413 &  13.77&   -350$^1$  & 200& R1 TNG  &-181  &  5 & R1 LOI &1.09 &E& B516     \\
2MASX-J01011110+3915047 &  13.44&   18100     &1500& R1 TNG  &      &    &       & & &    \\
2MASX-J01014341+4024207 &  13.53&   24900     & 500& R1 TNG  &      &    &       & & &    \\
2MASX-J00272474+4507360 &  14.23& M Star      & 38 & R2 TNG  &      &    &       & & &    \\
2MASX-J00312909+4053553 &  13.22&   20765     &  22& R1 LOI  &      &    &       & & &    \\
2MASX-J00382837+4608217 &  13.48&   -22       &  21& R1 LOI  &-18   &  9 & R2 LOI &1.00 &PS&    \\
2MASX-J00290822+3928357 &  12.41&   10887(EL) &  18& R1 LOI  &      &    &       & & &   \\
2MASX-J00291911+3825166 &  12.98&   5830      &15  & R1 LOI  &      &    &       & & &   \\
2MASX-J00240694+4459072 &  14.24&   45340$^2$ &24  & R1 LOI  &      &     &R2 LOI & & &   \\
2MASX-J00515568+4707229 &  13.53&   14491     &16  & R2 LOI  &      &    &       & & &   \\
2MASX-J00271402+3548121 &  13.17&   18667     &24  & R2 LOI  &      &    &       & & &   \\
2MASX-J00422528+3742110 &  13.58&   11520     &57  & R3 LOI  &      &    &       & & &   \\
2MASX-J00595989+4154068 &  14.16&   -272      &54  & R2 TNG  &      &    &       &1.33 &E& B517   \\
2MASX-J01002390+4225505 &  14.02&   31228     &45  & R2 TNG  &      &    &       & & &    \\
2MASX-J01003776+4213045 &  14.00&   32105     &84  & R2 TNG  &      &    &       & & &   \\
2MASX-J00252922+3750258 &  14.34&   29579     &73  & R2 TNG  &      &    &       & & &   \\
2MASX-J00273099+4008015 &  14.18&   21384     &71  & R2 TNG  &      &    &       & & &   \\
2MASX-J00294944+4143240 &  14.07& M Star      &24  & R2 TNG  &      &    &       & & &   \\
2MASX-J00364114+3734534 &  14.43&   46257     &21  & R2 TNG  &      &    &       & & &   \\
2MASX-J00370125+3738403 &  14.47&   54659     &31  & R2 TNG  &      &    &       & & &   \\
2MASX-J00373023+3626149 &  13.04&   15875     &45  & R2 TNG  &      &    &       & & &   \\
2MASX-J00254284+4125357 &  13.51&   44726     &85  & R2 TNG  &      &    &       & & &   \\
2MASX-J00262799+4416097 &  14.18&   48977     &111 & R2 TNG  &      &    &       & & &   \\
2MASX-J00355252+3751415 &  14.06&   28412     &68  & R2 TNG  &      &    &       & & &   \\
2MASX-J00423531+3731294 &  13.98&   38784     &72  & R2 TNG  &      &    &       & & &   \\
2MASX-J00425347+3909135 &  14.21&   52861     &121 & R2 TNG  &      &    &       & & &   \\
2MASX-J00543871+4507287 &  13.88&   25418     &54  & R2 TNG  &      &    &       & & &   \\
2MASX-J00565026+4008522 &  14.07&   28742     &68  & R2 TNG  &      &    &       & & &   \\
2MASX-J01035016+4320352 &  14.31&   63234     &34  & R2 TNG  &      &    &       & & &   \\
2MASX-J00145479+3905041 &  14.40&   -200      &48  & R2 TNG  &      &    &       &1.06 &E& B518  \\
2MASX-J00162141+3412002 &  14.05&   51033     &154 & R2 TNG  &      &    &       & & &   \\
2MASX-J00213516+4813322 &  14.02&   24782     &24  & R2 TNG  &      &    &       & & &   \\
2MASX-J00513014+3407389 &  13.99&   -268      &47  & R2 TNG  &      &    &       &1.36 &E& B519  \\
2MASX-J01100019+4024210 &  13.77&   17236     &58  & R2 TNG  &      &    &       & & &   \\
2MASX-J00020201+5136242 &  13.86&   24441     &13  & R4 LOI  &      &    &       & & &   \\
2MASX-J00061191+4130134 &  13.17&   19054     & 7  & R4 LOI  &      &    &       & & &   \\
2MASS-J00504245+3254587 &  13.37&    -312     & 17 & R1 LOI  &      &    &       & & &Martin-GC1 \\
\hline
\end{tabular}
\end{footnotesize}
}
{
\begin{flushleft}
$^a$ Object with prominent emission lines in his spectrum.
$^b$ Object with M type spectrum (prominent TiO bands).\\
R1, R2, R3 etc., refer to the various observing runs at the TNG and
Cassini (Loiano) telescopes, see Sect.~2. The 2MASS counterpart of
the remote cluster discovered by Martin et al. (2006), here named
Martin-GC1, is in 2MASS-PSC instead of 2MASS-XSC as
all other entries. Mackey-GC1 is GC1 from Mackey et al. (2007).\\
$^1$Very uncertain RV estimates obtained during Run 1 at TNG;
superseded by the estimates from other
sources.\\
$^2$ The radial velocity has been estimated from the stacked
spectrum obtained by co-adding the spectra acquired in Run 1 and Run
2 at the Cassini telescope. 
\end{flushleft}
}
\end{center}
\end{table*}

\subsection{Radial velocities}

The heliocentric radial velocities ($V_r$) of the candidate globular clusters
were obtained by cross-correlation (CC) with the template spectra,
using the IRAF/fxcor package (see Tonry \& Davis  \cite{tonry} for
details of the technique).
We applied a square filter to dampen the highest and lowest frequency Fourier
components; if unfiltered, these frequencies produced broad features that
masked the narrow CC peaks. We fit the CC peaks with Gaussians.
The typical internal velocity errors on a single measure were $\sim 50$ km/s
for BFOSC and  $\sim 65$ km/s for DoLoRes spectra.

In both cases we averaged the RV measures obtained by the
CC with different template spectra and we adopted the
standard deviation of the different estimates as our final error on
RV, as done in G06a. This procedure allows a significant reduction
of the uncertainty on RV, in particular for the BFOSC spectra, that
were cross-correlated with five very-high-S/N spectra from our {\em
Master RV Templates} set (see Appendix A, in online edition). 
Note that all our
CC RV estimates have been independently checked by measuring the
average red/blue-shift on the strongest lines identified in each
spectrum, using standard IRAF/rv tasks. The RVs obtained with the
two independent methods are fully consistent, within the observing
errors.

\subsection{Optical Imaging}

To gain insight on the morphology of the most promising candidates and, in
particular, to discriminate between truly Extended sources and actual stars
(Point Sources) we obtained BFOSC optical images in white light for eight of
the CGCs listed in Tab.~\ref{oss} (those having the R (9) and E/PS (10) columns filled),
as done in G06a. The observations were performed during Run 2, 3 and 4.

In imaging mode,
BFOSC has a pixel scale of $0.58\arcsec/px$ and a total field of view of
$13.0\arcmin \times 12.6\arcmin$.
The exposure times ranged between 1 and 5 minutes, depending on
the brightness of the target and on the atmospheric conditions. The nights were
not photometric but clear; the typical seeing was around
1.5$\arcsec$-2$\arcsec$ FWHM.

The images have been corrected for bias and flat field with standard IRAF
procedures. Relative photometry,
FWHM and morphological parameters of each source in the frame - down to a 5
$\sigma$ threshold - was obtained with Sextractor
(Bertin \& Arnouts \cite{sex}).
Only non-saturated and isolated sources are retained in the final catalogs
(Sextractor quality flag =0).
The relatively wide field of view of BFOSC
allows the simultaneous imaging of the target and
several field stars, thus generally allowing a robust discrimination
between Extended and Point Sources (see G06a and below).

   \begin{figure}
   \centering
   \includegraphics[width=9cm]{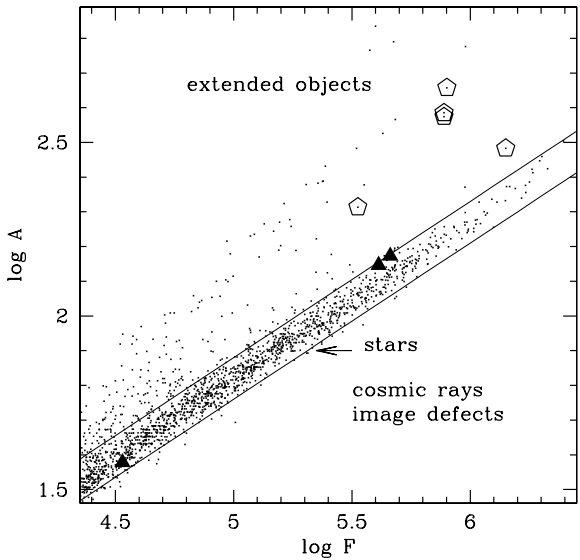}
    \caption{F-A diagrams from optical imaging for the considered targets
    (see G06a for discussion of the method).
    The pentagons are the CGCs classified as Extended objects
    (B516, B517, B518, B519
    and Mackey-GC1), filled triangles are the CGCs classified as Point Sources,
    i.e. 2MASX-J00285169+3905167, 2MASX-J00442168+3843319, and
    2MASX-J00382837+4608217. The parallel lines enclose the strip of the diagram
    that is populated by bona-fide stars (PS).
    The imaging of different fields have been renormalized to a unique scale by
    matching the sequence of stars.}
           \label{AF}
    \end{figure}
%
   \begin{figure}
   \centering
   \includegraphics[width=9cm]{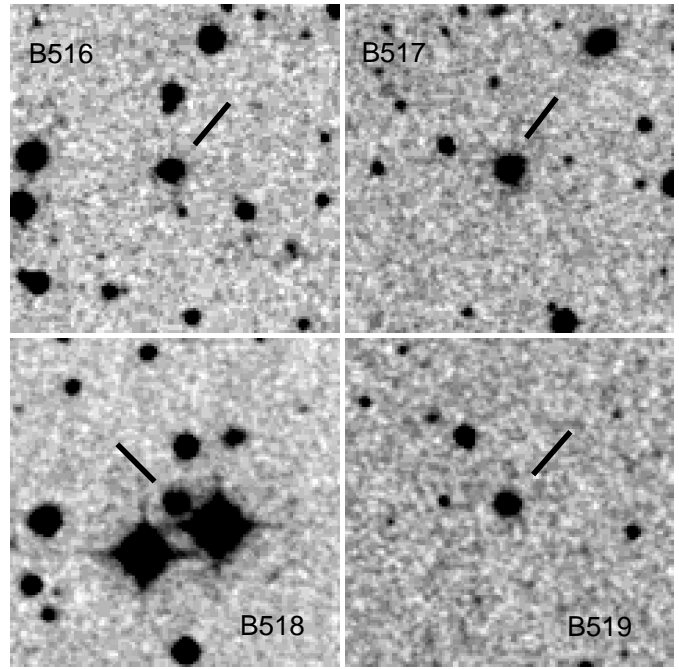}
    \caption{DSS2 images of the newly discovered clusters. Each image is
$\simeq 2\arcmin$ on a side.
North is up, East is to the left. Note that B518 lies between two very
bright foreground stars.}
           \label{image}
    \end{figure}

\section{Classification of candidates GC}

The essential results of the survey are reported in Tab.~\ref{oss},
that lists the 2MASS designation of the surveyed CGCs (name; col.
1), the K magnitude from 2MASS (transformed as in G04; col. 2), a
first estimate of RV, the associated uncertainties and the
observing run during which the  spectra has been acquired (cols.
3, 4, and 5, respectively), a second - typically more accurate - RV
estimate, the associated uncertainty and the observing run (cols. 6,
7, and 8, respectively), the ratio R between the FWHM of the target
and the FWHM of stars as measured on optical images (col. 9), a
morphological classification based on the Flux-Area diagram (E =
Extended, PS = Point Source, col. 10; see G06a and below), and the
name assigned here or by other authors to the candidates that has
been confirmed as bona-fide GCs (col. 11). The RV estimates of
column 6 supersedes those presented in col. 3, when both are present
(see above). Note that the 2MASS designation provides also the J2000
equatorial coordinates of the object: for example the coordinates of
2MASX-J00255466+4057060 are RA = 00$^h$ 25$^m$ 54.7$^s$ and Dec =
+40$\degr$ 57$\arcmin$ 06.0$\arcsec$.

Most of the classification work is quite straightforward:
Tab.~\ref{oss} contains 35 sources that have large recession
velocity ($V_r>+5000$ km/s), clearly of cosmologic origin. Hence
they are distant background galaxies (class 4 in the Revised Bologna
Catalogue - RBC - see G06a, and references therein \footnote{ 
see the Appendix B (published
only in the electronic edition of the Journal).}).

Five of the remaining 12 objects display strong TiO bands
typical of M stars in their spectra. The spectra of our RV templates are
clearly too different from those of these sources for a safe application of the
CC technique. For this reason we don't provide RV estimates for these sources.
However, since genuine Globular Clusters have spectral types earlier than K0
(Harris \cite{h96}), they clearly cannot be M31 GCs, and we classify all of
them as (likely foreground) stars (RBC class 6).

Of the remaining eight candidates, {\em five} have $V_r<-150$ km/s, hence they
may be classified as genuine M31 GCs based on their radial velocity alone,
according  to the criteria by G06a. The other three have RV typical of the
Galactic foreground population ($\langle V_r\rangle = -29$ km/s and
$\sigma=42.6$ km/s, G06a) but still compatible with M31 GCs. Their spectral
type is compatible with being GCs, hence we must recur to morphological
criteria to ascertain their nature.
From the optical imaging described in Sect.~2.4 above
we derived the ratio R between the FWHM of the candidate and that of
stars in the same image. In general
extended objects should have $R > 1.0$. We find here that all the considered
CGCs having $-150$ km/s $<V_r<$+100 km/s have $R\simeq 1.0$,
typical of point sources.
Moreover, in a Flux-Area diagram (F-A, see G06a) they lie in the characteristic
strip of Point Sources (see Fig.~\ref{AF}). Hence we classify them as
foreground stars (RBC class 6). It is somehow unexpected that our sample -
selected from a catalogue of extended sources (XSC) - contains also stars (see
Sect.~1.1, above).
This is probably due to the modest spatial resolution of the original 2MASS
images that may blend one or more stars into a spurious
extended source.

All the five candidates with $V_r<-150$ km/s have $R\ge 1.06$, up to
$R= 1.36$, and appear as  Extended sources in the F-A diagram of
Fig.~\ref{AF}. Hence we classify all of them as genuine M31 GCs
(RBC class 1, but see Sect.~3.2, below).
It turned out that 2MASX-J00264769+3944463 was already
(independently) identified by H04; it is listed as
GC1 in Mackey et al. \cite{mackhux1}. The HST imaging by these authors confirms
that this is a genuine M31 globular.
In the following we will refer to this
cluster as Mackey-GC1, in analogy to the case of
Martin-GC1\footnote{These are provisional names. A. Huxor and
collaborators will assess the nomenclature of the clusters they discovered in
the INT-WFC and MEGACAM surveys in a paper that is in preparation (A. Huxor,
private communication). In the current version of the RBC we have adopted the
following abbreviated names for these clusters: Mackey-GC1 = MCGC1,
Martin-GC1 = MGC1.}, while the assigned RBC name is B520 (see below).

On the other hand, 2MASX-J00553861+4524413, 2MASX-J00595989+4154068,
2MASX-J00145479+3905041, 2MASX-J00513014+3407389, have never been recognized
before as M31 globular clusters. They are completely new discoveries and we
christen them B516, B517, B518, and B519\footnote{The name B515 was assigned
to a cluster recently discovered on HST images, as reported
by G06a.}, respectively, according to the RBC
nomenclature. These new discoveries increase the number of known
remote M31 clusters
($R_p\ge 40$ kpc) from five (B514/Mackey-GC4, Mackey-GC1/B520, Mackey-GC5,
Mackey-GC10 and Martin-GC1) to nine clusters. If we consider clusters having
$R_p\ge 35$ kpc the sample increases from 8 to 12 clusters
(see Fig.~\ref{RMv}, below).

\begin{table}
\begin{center}

\caption{Remote clusters: positions and magnitudes \label{clus}}
{
\begin{scriptsize}
\begin{tabular}{l c r r r r r r r}
\hline \hline
Name       & V& $r\prime$ &X$\prime$& Y$\prime$& R$\prime$&   R   & $V_r$&$ \pm\epsilon_{Vr}$ \\
           & mag &  mag$^1$ &         &          &          &  Kpc  &    km/s &  km/s   \\
\hline
Mackey-GC1 &  15.0&16.18 &-181.6 &  92.2 & 203.7 & 46.4& -219  & 15  \\
B516       &  15.3&15.55 & 282.5 &  46.2 & 286.2 & 65.2& -181  &  5  \\
B517       &  16.1&16.13 & 151.8 &-126.3 & 197.5 & 45.0& -272  & 54   \\
B518       &  16.0&16.27 &-292.6 & 184.3 & 345.8 & 78.8& -200  & 48   \\
B519       &  17.3& --   &-272.7 &-349.5 & 443.3 &101.0& -268  & 47   \\
Martin-GC1 &  15.5&15.28 &-336.4 &-338.4 & 513.9 &117.0& -312  & 17  \\
\hline
\end{tabular}
\end{scriptsize}
}
{
\begin{flushleft}
$^1$ Homogeneously calibrated CCD magnitudes in the Sloan Digital
Sky Survey $r^{'}$ passband from the Carlsberg Meridian Catalogue 14
{\tt http://www.ast.cam.ac.uk/~dwe/SRF/cmc14.html}. 
\end{flushleft}
}
\end{center}
\end{table}

DSS2 post stamp images of the new clusters are presented in
Fig.~\ref{image}. It is interesting to note that B518 and Martin-GC1 are listed
in the Hyperleda database of galaxies (Paturel et al. \cite{leda}); hence they
have been independently recognized as extended sources also in that catalogue.
Among the targets with non-cosmological velocities, the only other source
included in Hyperleda is 2MASX-J00442168+3843319 - that we clearly identify as a
point  source. In support to our conclusion, the NOMAD catalogue (Zacharias et
al. \cite{nomad}) report a total proper motion of 168 mas/yr for this source;
all the targets we classified as clusters have estimated proper motions $<10\pm
10$ mas/yr (i.e. consistent with zero, if any), as expected for
extra-galactic sources.

   \begin{figure}
   \centering
   \includegraphics[width=9cm]{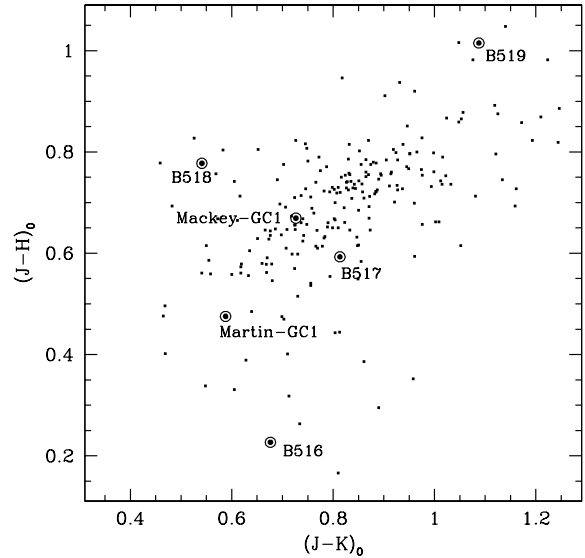}
    \caption{De-reddened near infrared colors of confirmed M31 GCs from the RBC
    (Galleti et al. \cite{silvia}; small points).
    A reddening of E(B-V)=0.1 has been assumed for all the clusters.
    The remote clusters identified in our survey are plotted as larger symbols
    (encircled filled circles) and labeled. Their reddening has been taken from
    the COBE/DIRBE maps by Schlegel, Finkbeiner \& Davis \cite{sfd}.}
           \label{colori}
    \end{figure}

\subsection{The physical characteristics of the new clusters}

Some useful parameters of the remote clusters studied here are summarized in
Tab.~\ref{clus}.
The V magnitudes have been drawn from the NOMAD catalogue (Zacharias et
al. \cite{nomad}) and from M06b for Martin-GC1.

All the considered clusters are more than $3\degr$ away from
the center of M31. With the adopted distance scale, the projected distance of
these clusters goes from $R_p=45$ kpc to $R=101$ kpc. Martin-GC1
($R_p=117$ kpc) remains the outermost known GC of M31, to date. Note that RGB
stars belonging to the halo of M31 have been traced out to $\sim 165$ kpc from
the center of M31 (Kalirai et al. \cite{kali}).

Near Infra Red (NIR) colors of the considered clusters\footnote{The
colors of remote clusters has been corrected for extinction using
the COBE-DiRBE reddening maps by Schlegel et al. \cite{sfd}. The
disc of M31 is a non-subtracted source in the COBE-DiRBE maps; for
this reason the reddening derived from the maps is - in general -
incorrect for sources projected unto the disc itself. The remote
clusters considered here are too far from the disc of M31 to be
affected from this kind of problem, hence reliable estimates of the
average foreground  extinction can be obtained from Schlegel et
al.'s maps. For the other RBC clusters plotted in Fig.~\ref{colori}
we adopted $E(B-V)=0.1$, as in G06b.} are compared to those of the
known confirmed M31 GCs from the RBC in Fig.~\ref{colori}. All the
newly discovered clusters have colors within the range covered by
known confirmed M31 globular clusters. B516 appears very blue in
$(H-K)_0$ and B519 appears as one of the reddest M31 GCs. These
possibly odd positions in the NIR color-color diagram may be due to
(a) random fluctuations within the sizable uncertainties affecting
the photometry, typically of order $\simeq 0.05 - 0.1$ mag for these
clusters, (b) local variations of the interstellar extinction,
and/or (c) intrinsic differences in physical parameters (age,
metallicity).

\begin{figure}
   \centering
   \includegraphics[width=9cm]{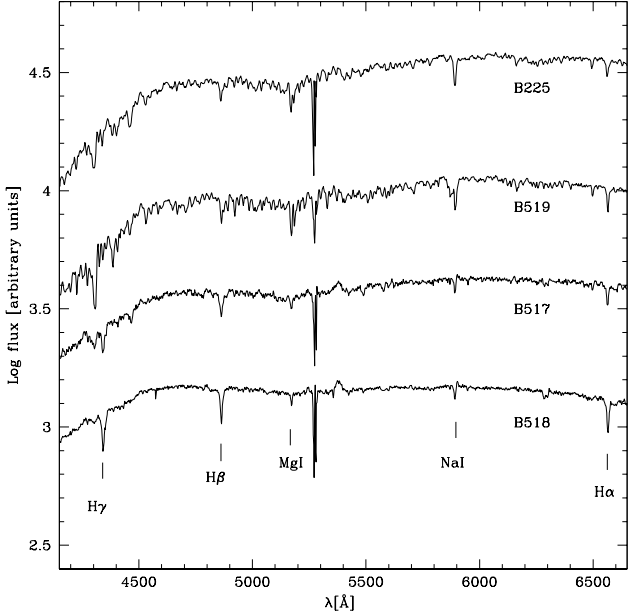}
    \caption{DOLORES spectra of target clusters. From top to bottom, the RV
    template B225, a well known metal-rich cluster, reported here for
    comparison, B519, B517, and
    B518. The main spectral features have been labeled.
    The strong feature at $\simeq 5250\AA$ is a spurious line due to
    a defect of the CCD.
    Note the prominent Mg lines in the spectrum of B519, suggestive of a metal
    content similar or larger than that of B225 ([Fe/H]$\sim -0.45$).}
           \label{spetng}
    \end{figure}

\begin{figure}
   \centering
   \includegraphics[width=9cm]{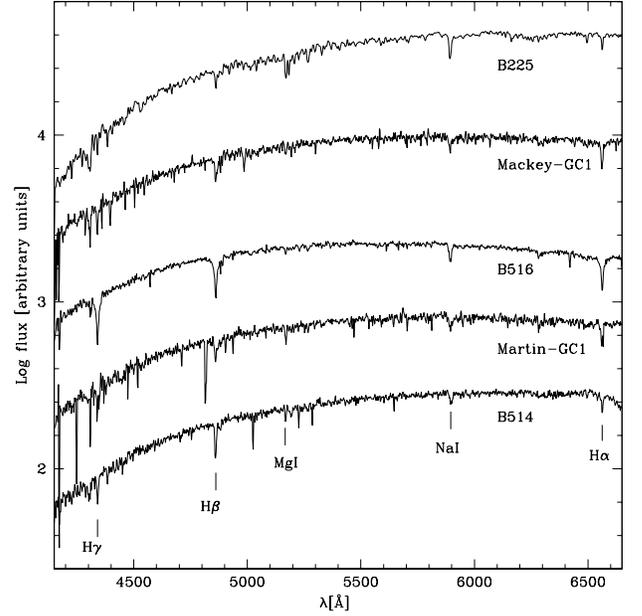}
    \caption{BFOSC spectra of target clusters. From top to bottom, the RV
    template B225, a well known metal-rich cluster, reported here for
    comparison, Mackey-GC1, B516,
    Martin-GC1, and B514, a remote old and metal poor cluster (G05, G06b).
    The main spectral features have been labeled.
    Note the prominent Balmer's lines in the spectrum of B516.}
           \label{speloi}
    \end{figure}

A further insight in can be obtained by inspection of the spectra,
shown in Fig.~\ref{spetng} (DoLoRes spectra) and
Fig.~\ref{speloi} (BFOSC spectra).
In both figures we report
also the spectrum of a well known and bright M31 GC, the metal rich cluster
B225 ($[Fe/H]\simeq -0.45$, Puzia et al. \cite{puzia})\footnote{ Puzia et
al. \cite{puzia} report all metallicities in [Z/H] dex. A transformation to [Fe/H]
 has been done through the equation: [Fe/H]= [Z/H] - 0.94 [$\alpha$/Fe] from
Thomas et al. \cite{thomas}.},
observed with the same instrument,
for comparison.

Fig.~\ref{spetng} shows that B517 and B518 have weaker MgI and NaI lines and
stronger Balmer lines with respect to B225. This suggest that they may be old
globulars, significantly more metal deficient than B225 (see below).
On the other hand, B519  has Balmer lines very similar to B225 and metal lines
as strong or stronger than B225, thus suggesting that this may be a
quite metal rich globular cluster.

In Fig.~\ref{speloi} we plotted also the spectrum of the remote old and
metal-poor cluster B514 ($[Fe/H]\simeq -1.8$, G05, G06b).
Mackey-GC1 and Martin-GC1 have MgI and NaI lines much weaker than B225; their
overall spectrum is very similar to that of B514, i.e. of an old and metal poor
globular with a Blue Horizontal Branch (BHB). This is in full agreement with the
results obtained by M06b and by
Mackey et al. \cite{mackhux1} from the CMD of the clusters.
On the other hand B516 shows strong NaI and weak MgI features, while Balmer's
lines are much stronger that those observed in the old metal-poor  and BHB
clusters. These characteristics suggests a significantly younger age for this
cluster.

%
\begin{table*}
\begin{center}

\caption{Remote clusters and template clusters: Lick indices and
metallicities \label{Lick}}
{
\begin{tabular}{l c c c c c}
\hline \hline
name       & $H_{\beta}$ & Mg$_2$ & Mgb   & [Fe/H]$_{Mg2} $ & [Fe/H]$_{Mg2} $\\
           &  [$\AA$]    &[mag]&[$\AA$]&   Ea.~1         & BGM92$^1$     \\
\hline
Mackey-GC1 &1.841 $\pm0.286$ &0.023 $\pm 0.007$ &0.688 $\pm 0.290$ &  -2.0&  -1.9 \\
B516       &4.818 $\pm0.146$ &0.011 $\pm 0.004$ &0.684 $\pm 0.165$ &   -- &  -- \\
B517       &2.567 $\pm0.128$ &0.089 $\pm 0.004$ &1.066 $\pm 0.150$ &  -1.3&  -1.4 \\
B518       &3.367 $\pm0.088$ &0.057 $\pm 0.002$ &1.000 $\pm 0.106$ &  -1.6&  -1.6 \\
B519       &1.996 $\pm0.026$ &0.195 $\pm 0.001$ &3.592 $\pm 0.028$ &  -0.4&  -0.6 \\
Martin-GC1 &1.595 $\pm0.283$ &0.020 $\pm 0.007$ &1.012 $\pm 0.285$ &  -2.1&  -1.9\\
\hline
B158 Loi  &1.853 $\pm0.092$ &0.112 $\pm 0.002$  &2.390 $\pm 0.090$ &  -1.1& -1.2 \\
G001 Loi  &2.239 $\pm0.103$ &0.110 $\pm 0.003$  &2.301 $\pm 0.104$ &  -1.1& -1.2 \\
B514 Loi  &2.317 $\pm0.132$ &0.044 $\pm 0.003$  &0.422 $\pm 0.137$ &  -1.8& -1.7 \\
B225 Loi  &1.725 $\pm0.043$ &0.167 $\pm 0.001$  &3.449 $\pm 0.042$ &  -0.6& -0.8\\
B225 TNG  &1.775 $\pm0.062$ &0.204 $\pm 0.002$  &3.163 $\pm 0.068$ &  -0.3& -0.6\\
\hline
\end{tabular}
}
{
\begin{flushleft}
Below the horizontal line we list the data for already known
clusters, for
reference. \\
$^1$ Metallicity from the calibration provided by Buzzoni, Gariboldi
\& Mantegazza \cite{buzz}, [Fe/H]=7.41Mg$_2$-2.07.
\end{flushleft}
}
\end{center}
\end{table*}

As for B514, we measured the strongest Lick's indices that are
accessible within our spectra (Mg$_2$, Mgb, $H_{\beta}$),
following the procedure illustrated in G05 and adopting the definitions by
Trager et al. \cite{trager}. The comparison between the indices measured by us
and those reported by Trager et al.'s for the clusters in common
(B225, B158, see Tab.~\ref{Lick}, below)
suggests that our indices are in reasonable agreement with the original Lick
scale; however they should be considered as preliminary estimates until a more
detailed comparison allows us to apply a robust transformation from our natural
systems to the
actual Lick system (Galleti et al. 2007, in preparation; hereafter G07).
The measured indices are reported in Tab.~\ref{Lick}. We use the following relation
calibrated on Galactic globulars (G07), to obtain spectroscopic estimates of
the metallicity from the Mg$_2$ index

\begin{equation}
[Fe/H]=-2.32+12.89Mg_2-15.9Mg_2^2 ~~~~~rms=0.17 ~dex
\end{equation}

\begin{figure}
   \centering
   \includegraphics[bb= 100 214 550 550, width=9cm]{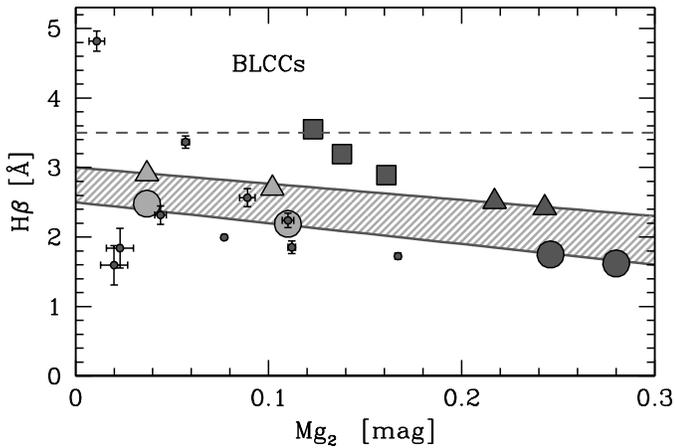}
    \caption{The $H_{\beta}$ and $Mg_2$ indices of the observed clusters (Tab.~3,
    small filled circles with error bars)
    are compared with the predictions of the models by Buzzoni et al.
    \cite{buzz}. The large circles correspond to the model of a Simple Stellar
    Population (SSP) of age 15 Gyr and metallicity [Fe/H]=--2.27, --1.27, --0.25,
    0.0, from left to right, with red Horizontal Branch morphology.
    The triangles correspond to the same model at the same metallicities but
    with an intermediate HB morphology, instead. Therefore, the shaded area
    encloses the region of the plane that is expected to be populated by old
    globular clusters. The filled squares correspond to a model of age 2 Gyr,
    with metallicity [Fe/H]=--0.25, 0.0, +0.3, from left to right respectively.
    The realm of the Blue Luminous Compact Clusters (Fusi Pecci et al. 2005) is
    marked by a long dashed line and labeled. The cluster with the highest
    $H_{\beta}$ is B516 and the cluster with the highest $Mg_2$ is B519.}
           \label{hbmg2}
    \end{figure}

The resulting metallicities are reported in Tab.~\ref{Lick}; they are in good
agreement with those obtained from the theoretical calibration by
Buzzoni et al. \cite{buzz}, that are also reported in the same table,
for comparison.
The estimates for Mackey-GC1
and Martin-GC1 are in agreement with those presented in Mackey et al.
\cite{mackhux1} and in M06b, respectively
(i.e. $[Fe/H]\sim -2.0$). B517 and B518 seem slightly less metal deficient
($[Fe/H]\sim -1.3$ and $\sim -1.6$, respectively. B519 clearly emerges as a
metallicity outlier among  the remote clusters, having $[Fe/H]\simeq -0.4$.
The adopted $[Fe/H]$ vs Mg$_2$ relation is valid only for classical old globular
cluster. The strong Balmer lines suggest that B516 may be much younger
than classical globulars, hence the derived metallicity estimate is
probably unreliable (consequently, it is not reported in Tab.~\ref{Lick}).
On the other
hand, the measured $H_{\beta}>4.5\AA$ includes B516 among the Blue Luminous
Compact Clusters (BLCC) as defined by Fusi Pecci et al. \cite{ffp}. The
integrated spectra and color of these clusters indicate ages $\le 2$ Gyr, as can
be appreciated from the comparison with Buzzoni et al. \cite{buzz} models shown
in Fig.~\ref{hbmg2}.

\subsection{B516 and B519: supplementary investigation}

As said above, B516 and B519 fulfil all the criteria devised by G06a to
classify a CGC as a genuine M31 star cluster. However both present some
peculiar property, requiring some supplementary investigation on
their nature.

As said,  B519 is  the only
one of our newly discovered clusters that lies within the boundaries of the
MEGACAM survey (M06b). N.F. Martin kindly
inspected the images from their deep optical survey and provided us with an
image of the target and its surroundings. B519 clearly appears as an extended
source, with a halo of diffuse light; the MEGACAM image fully confirms that it
cannot be neither a star or a blend of two or more stars. Unfortunately, the
image is not partially  resolved into stars as in the case of Martin-GC1; from
the mere inspection of the images one would have classified B519 as an
elliptical galaxy.
However the  non-resolved nature of the image is not
conclusive, since there are cases in which genuine M31 GCs cannot be (even
partially) resolved into stars even with HST imagery (see, for example, Barmby
et al. \cite{bar07}).

In the scenario outlined above, the final word is left to the radial
velocity estimate. The estimate presented in Tab.~\ref{oss} is
obtained from a very clean, single and strong cross-correlation peak
($CC\simeq 0.4$ and Tonry \& Davis' parameter $TDR\simeq 20$). We
obtained an alternative estimate of RV from the same spectra by
measuring the wavelength shift (with respect to restframe) of seven
strong lines (including Balmer lines): the average velocity is
$\langle V_r \rangle = -241$ km/s, the standard deviation
$\sigma=36$ km/s, in full agreement with the CC result. Since (a) it
is out of doubt that B519 is a {\em bona fide} extended object, and (b) it
has a velocity typical of M31 GCs and incompatible with cosmological
recession, {\em we firmly conclude that B519 is a genuine remote - and metal
rich - star cluster of M31}.

The RV estimate rules out a background galaxy as a viable hypothesis
also for B516. Unfortunately, in this case we lack imaging of
sufficient quality to rule out the possibility that it is a spurious
extended object due to the superposition of two or more unrelated
stars. The R parameter is not extreme (R=1.09). In Fig.~\ref{AF},
B516 is the open pentagon with the highest value of log F: the point
is outside the ``star strip'', but not so clearly on the ``extended
object'' branch of the diagram as the other clusters studied here.
Cohen, Mattews \& Cameron \cite{coh} have demonstrated that some
objects classified as M31 BLCC because of their strong $H_{\beta}$
and their RV typical of M31 cluster are in fact asterisms, i.e.
chance superpositions of stars residing in the star-forming thin
disc of M31.  This kind of spurious clusters can be encountered only
in the high-surface brightness inner regions of the M31 disc, where
the strong stellar crowding makes relatively likely the possibility
of finding several bright M31 stars within a very small angle( $\la
5 \arcmin$).  This cannot be the case of B516, that lies at $\sim
4.8\degr$ ($R_p\simeq 65$ kpc) from the center of the galaxy, where
no sign of the disc of M31 is visible. It remains the possibility
that B516 is a (or a blend of) Galactic F or A  spectral type
star(s). Such star should likely be a member of the Galactic Thin
Disc since these early types are very rare in the Thick Disc and in
the Halo. The synthetic sample of Galactic stars in the foreground
of M31 that we extracted from the Robin et al. \cite{robin} model
and we analyzed in G06a does not contain any Thin Disc star having
$V_r<-150$ km/s; actually, only 1\% of the whole sample have
$V_r<-150$ km/s. In the same synthetic sample, Thick Disc stars have
$\langle V_r \rangle \simeq -53$ km/s and dispersion $\sigma\simeq
54$ km/s; the rare Halo stars have $\langle V_r \rangle \simeq -196$
km/s and  $\sigma\simeq 96$ km/s, hence the RV of B516 is not
strictly incompatible with these Galactic components.
The Monoceros Ring substructure (Yanny et al. \cite{yanny}; Ibata et al.
\cite{ibaring}) crosses the line of sight to M31 and includes A, F stars;
however Martin et al. \cite{monoceros} have found that Monoceros Ring stars in
this direction have mean velocity $\langle V_r \rangle \simeq -75$ km/s and
dispersion  $\sigma\simeq 26$ km/s, which seem incompatible with the velocity
of B516.

Given the scenario described above, a field Blue Straggler belonging to the
Galactic Halo (see Carney et al. \cite{carney}, and references therein) may be
the more likely alternative explanation for the nature of B516.

In conclusion, the formal application of the criteria adopted in
G06a would lead to classify B516 as a genuine M31 GC; if this were
the case we should also conclude that B516 is a young bright cluster
as those described by Fusi Pecci et al. \cite{ffp}.  The occurrence
of a young cluster so far away from the regions of M31 where the
star formation is presently ongoing is unexpected; since neither the
RV, nor the available imaging can completely exclude the possibility
that B516 is a  foreground source we maintain a doubt on the final
classification of this object, until further observations will
clarify the issue.

\section{The population of remote GCs of M31}

\begin{figure}
   \centering
   \includegraphics[width=9cm]{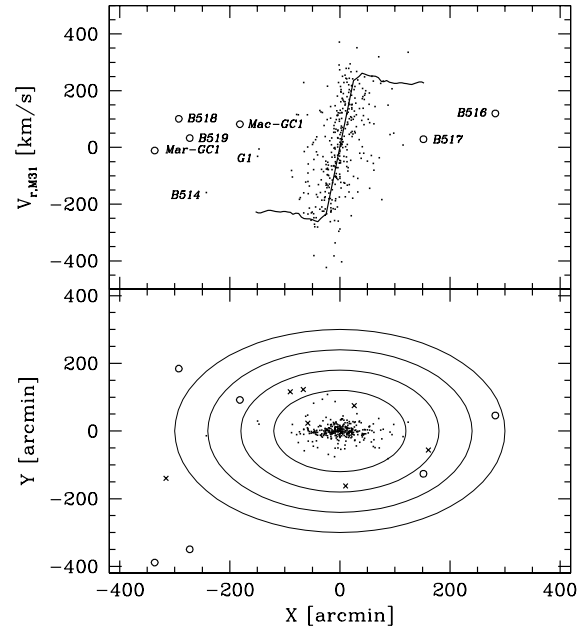}
    \caption{Upper panel: velocity distribution along the major axis for
    already confirmed M31 GCs (small points), and for remote clusters studied in
    the present paper (open circles). The line is the HI rotation curve from
    Carignan et al. \cite{cari}. $V_{r,M31}$ is the line-of-sight velocity in
    the Andromeda-centric reference system.
    Lower panel: position of the clusters in the sky in the Andromeda-centric
    reference system; open circles represent remote clusters discovered in
    our survey, ``x'' symbols mark clusters discovered by other teams.
    The overplotted
    concentric {\em circles} have radius 2,3,4, and 5 degrees.}
           \label{vrot}
    \end{figure}

In Fig.~\ref{vrot} we show the velocity (upper panel) and spatial
(lower panel) distribution of the confirmed M31 GC from the RBC and
of the recently discovered remote clusters. The HI rotation curve
from Carignan et al. \cite{cari} is also plotted in the upper panel,
as a reference. The sample of remote clusters currently available
for kinematical analysis does not show any clear correlation between
position and M31-centric velocity ($V_{r,M31}$). This seems
consistent with the association of these clusters to the
non-rotating very extended and metal-poor stellar halo that has been
recently found in M31 (Kalirai et al. \cite{kali}, Chapman et al.
\cite{chap}). The velocity dispersion computed from the six objects
listed in Tab.~\ref{clus} plus B514 is $\sigma =94\pm 46$ km/s, in
agreement with what found by Chapman et al. \cite{chap} for the
stellar halo ($\sigma\simeq 126$ km/s). A detailed analysis of the
kinematics of the M31 GC system as a whole is beyond the scope of
the present contribution and it is demanded to the end of our
survey: it is quite clear from Fig.~\ref{vrot} that the (presumably)
high incompleteness of the sample prevents any further speculation.

%

   \begin{figure}
   \centering
   \includegraphics[width=9cm]{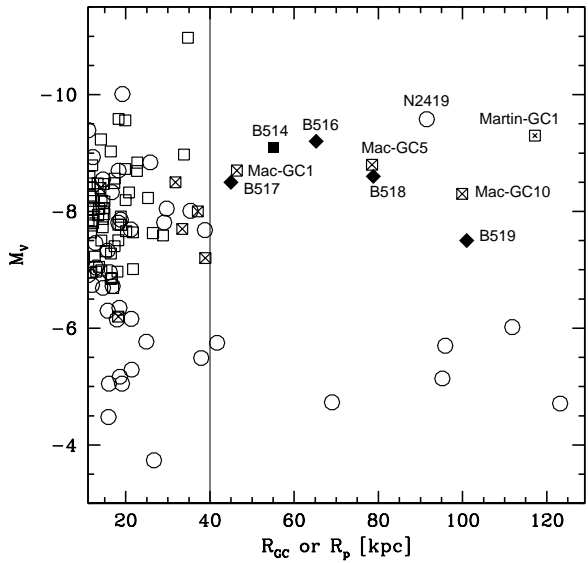}
    \caption{Absolute integrated V magnitude versus galactocentric distance (or {\em
projected} galactocentric distance) for MW (open circles) and M31 globular
clusters (open squares). The crossed squares are the new clusters
recently presented by Mackey et al. \cite{mackhux1};
the open square with the small cross is the remote
Martin et al.'s GC1; the filled square is B514 (G05, G06b, Federici et al.
\cite{fedb514}); the filled
diamonds are the newly identified remote clusters.
The bright clusters at $R_{GC}>40$ kpc have been labeled. The vertical
line marks $R_{GC/p}=40$ kpc.}
           \label{RMv}
    \end{figure}

At the present stage, the most interesting characteristic of our
(scanty) sample of remote M31 GCs emerges from the comparison with
the case of the Milky Way and is illustrated in Fig.~\ref{RMv} (see
also Mackey et al. \cite{mackhux1}). It is quite clear that the
outskirts of M31 host a significant population of bright clusters
($M_V\le-7.5$) whose only Galactic counterpart is NGC~2419, indeed a
quite peculiar cluster (see Mackey \& van den Bergh \cite{macksyd},
Federici et al. \cite{fedb514}, and references therein). The
joint luminosity distribution of remote clusters of both M31 and the
MW appears curiously bimodal, with a ``faint'' population ($M_V\sim
-5$) dominated by  MW clusters and a ``bright'' population ($M_V\sim
-8$), constituted by M31 clusters plus NGC2419, with a wide gap in
the middle ($M_V\sim -7$). Note that while ``faint'' clusters are
(probably) missing in the M31 sample for obvious reasons of incompleteness, the
lack of ``bright'' and intermediate luminosity clusters in the MW is
clearly a real effect. The difference between the remote GC
populations of the two galaxies is even more striking if one takes
into account that the M31 sample is clearly very far from complete
(see Sect.~1.1.1): several bright remote clusters are probably still
to be discovered.

At present, there is no obvious explanation for
the luminosity distribution of remote clusters shown in
Fig.~\ref{RMv}.
F07 pointed out that some of the newly discovered remote
M31 GCs have half-light radii that are too large for their
luminosity with respect to ordinary globulars, resembling the nuclei
of dwarf elliptical galaxies instead. Mackey \& van den Bergh
\cite{macksyd} noted that all the clusters having this
characteristic (as, for instance NGC2419) have been proposed in the
past, for various reason, as the possible remnants of nucleated
dwarf galaxies. If part of the bright remote clusters of
Fig.~\ref{RMv}
are the survived
nuclei of disrupted galaxies, the difference between the MW and M31
populations can be possibly interpreted in terms of initial
conditions, that is, for example, it may have been determined by a
difference in the luminosity function and/or in the distribution of
morphological types of the original building blocks from which the
two galaxies were assembled.

In any case, it is quite clear that the study of these remote clusters may
reveal us fundamental pieces of information about the early stage of the
formation of the Milky Way and M31.

\begin{acknowledgements}

We thank the referee, P. Barmby, for her useful and prompt report.
We are grateful to N.F. Martin and A.P. Huxor for their kind collaboration
in cross-checking our results.
A.B. acknowledges the support by INAF through the INAF/PRIN05 grants 1.06.08.02
and 1.06.08.03.
We are grateful to the staff of the
Telescopio Nazionale Galileo and Loiano Observatory for their kind
assistance during the observing runs. This research has made use of
the VizieR catalogue access tool, CDS, Strasbourg, France, and
of NASA's Astrophysics Data System.
The Digitized Sky Surveys were produced at the Space Telescope Science
Institute under U.S. Government grant NAG W-2166.
\end{acknowledgements}

\bibliographystyle{aa}

\newpage

\Online

\begin{appendix}

\section{Master RV templates for BFOSC}

In the past few years we had several observational runs dedicated to the
acquisition of BFOSC spectra of M31 clusters at the Cassini telescope. In all
the cases we adopted the same set up described here and during every night of
observation we acquired at least one spectrum of a bright well-known M31 cluster
for which a high-accuracy estimate of RV is available
(from Peterson \cite{pet} or Dubath \& Grillmair \cite{DG}, see G06a), to be
used as template for the cross correlation. In this way we have assembled a
conspicuous number of relatively high S/N spectra of template clusters. We have
verified that in all cases the spectra of the same template taken in different
runs gave null velocity difference once cross-correlated, hence the velocity
scale of BFOSC seems very stable. Given all the above, we decided to stack all
the spectra of a given template cluster (taken in different nights/runs)
to obtain master spectra with the
highest possible S/N ratio. The use of stacked spectra of a set of templates
would significantly improve the accuracy of the RV estimate of target clusters.

\begin{figure}
   \centering
   \includegraphics[width=9cm]{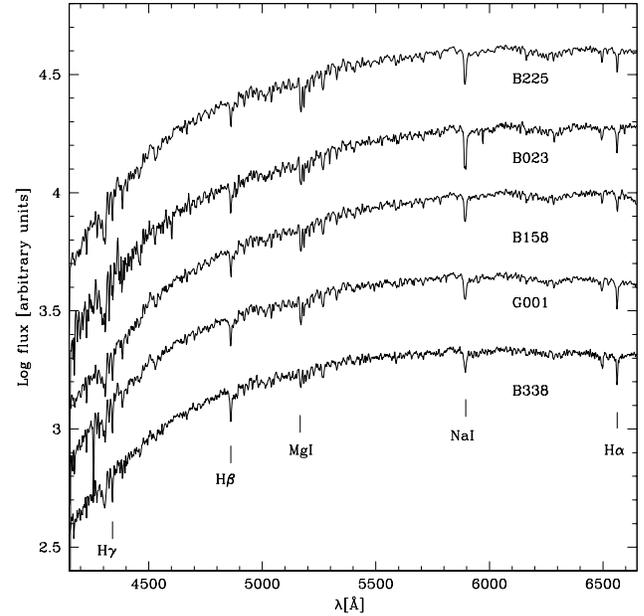}
    \caption{BFOSC spectra of template clusters. From top to bottom, the RV
    templates B225, B023, G001, B158, B338
    cluster (G05, G06b).
    The main spectral features have been labeled. The typical S/N of this
    spectra is $\sim 100$ for $\lambda > 4700\AA$.
    }
           \label{spetemp}
    \end{figure}

Here we present the list of {\em Master RV templates} that has been used for the
present study (Tab.~\ref{tabtemp}); the stacked spectra are shown in Fig.~\ref{spetemp}.

In future BFOSC runs we will acquire spectra of templates during each
observing night for the following purposes:

\begin{enumerate}

\item to verify that the acquired spectra are in the same velocity scale
      of the Master RV templates.

\item to obtain more spectra to stack into the Masters.

\item to include more clusters in the list of the Master RV templates.

\end{enumerate}

The final goal of the project is to assemble high S/N master spectra
for $\sim 10$ RV templates of various metallicities, to maximize the
return of future spectroscopic campaigns at the Cassini telescope.

%

\begin{table}
\begin{center}
\caption{Master RV template clusters (BFOSC) \label{tabtemp}} 
{
\begin{tabular}{l c r r c c}
\hline \hline
Name     & V&$V_r\pm\epsilon_{Vr}$& N(oss)&[Fe/H]& S/N\\
           & mag    &     km/s &  dex            &    \\
\hline
B023       &  14.22 & -451$\pm$ 5 &  2 &   --  & 140\\
B158       &  14.70 & -187$\pm$ 1 &  5 & -0.70 & 200\\
B225       &  14.15 & -165$\pm$ 1 & 13 & -0.45 & 500\\
B338       &  14.25 & -274$\pm$ 3 &  2 & -1.17 &  90\\
G001       &  13.21 & -332$\pm$ 3 &  3 & -1.02 & 115\\
\hline
\end{tabular}
}
{
\begin{flushleft}
Metallicities are taken from Puzia et al. \cite{puzia}. We have
transformed [Z/H] into [Fe/H] from the equation: [Fe/H]= [Z/H] -
0.94 [$\alpha$/Fe] by Thomas et al. \cite{thomas}
\end{flushleft}
}
\end{center}
\end{table}


\end{appendix}

\begin{appendix}

\section{Updates to the RBC}

As a duty cycle operation to maintain the RBC as updated as
possible, we search the literature for new classification and data
on M31 GCs and we periodically search the HST archive for
intentional or serendipitous images of M31 CGCs that can potentially
reveal the true nature of some candidate.

Here we report on the results of the careful visual inspection of
archival HST images of 164 objects already included in the RBC,
irrespectively of their current classification, as it may happen
that a very high resolution image supersedes even a spectroscopic
classification (see G06a). The results are summarized in
Tab.~\ref{hst} where we report: the name of the target (col. 1); the
spectroscopic classification, as defined in G04: C - cluster, G -
galaxy,  S -star, H - H~{\sc ii} region (col. 2); previous
classification via high resolution imaging, as defined in G04: C -
cluster, G - galaxy, S - star, H - H~{\sc ii} region, A - asterism
(col. 3); the classification obtained from the visual inspection of
HST images in the present work\footnote{Classifications followed by
a question mark indicates that a completely firm conclusion on the
nature of an object cannot be reached based on the considered image
alone. For example, the classification ``globular cluster''
indicates an object clearly resolved into stars, ``globular
cluster?'' indicates the case of a roundish extended object
resembling a cluster but NOT clearly resolved into stars.}(col. 4);
the final classification flag adopted in the new version of the RBC
(1: confirmed GC, 2: GC candidate, 3: controversial object, 4:
confirmed  galaxy, 5: confirmed H~{\sc ii} region, 6: confirmed
star, 7: asterism) (col. 5); the Proposal Id number (col. 6); the
instrument (col. 7); the passband of the considered image (col. 8);
the image name (col. 9); the exposure time (col. 10).

\begin{table*}
\begin{center}

\caption{Newly discovered globular clusters candidates \label{hstnew}}

\begin{tabular}{l c r r r r r r }
\hline \hline
Name& c& RA & Dec &   Camera&Filter&Dataset&Exptime \\
\hline  
B521  &  2 &00:41:41.67   &+40:52:01.41&ASC/WFC  &F606W & j96q06010 &3250\\
B522  &  2 &00:41:50.94   &+40:52:48.29&ASC/WFC  &F606W & j96q06010 &3250\\
B523  &  2 &00:42:46.28   &+41:18:32.41&ACS/WFC  &F435W & j8vp05010 &2200\\
B524  &  2 &00:42:55.68   &+41:03:11.22&ACS/WFC  &F775W & j8hoiiq8q & 624\\
\hline
\end{tabular}

\end{center}
\end{table*}


Most of the inspected objects (130) were already classified in the RBC
(as star, cluster, galaxy etc), but only 43 objects were previously classified
with high resolution imaging.
34 objects were just CGCs that are
classified here for the first time.
Two of these candidates turn out to be genuine globular clusters (B202 and
AU010), B004D and B328D are clearly background galaxies, while all the other
candidates are stars or blended groups of stars or asterisms.

The controversial candidates B140 and B366 are clearly resolved into
stars in the inspected images, hence they are genuine clusters,
while B175 is clearly a star. B253 and B034D, classified as genuine
globular clusters by G06a based on their radial velocity are
recognized as stars in the considered images. B147 was classified as
a star by Dubath $\&$ Grillmair \cite{DG}, based on its velocity
dispersion: the HST images unambiguously reveal that this is a well
resolved star cluster, as recently pointed out also by Barmby
et al. \cite{bar07}.

   \begin{figure}
   \centering
   \includegraphics[width=8cm]{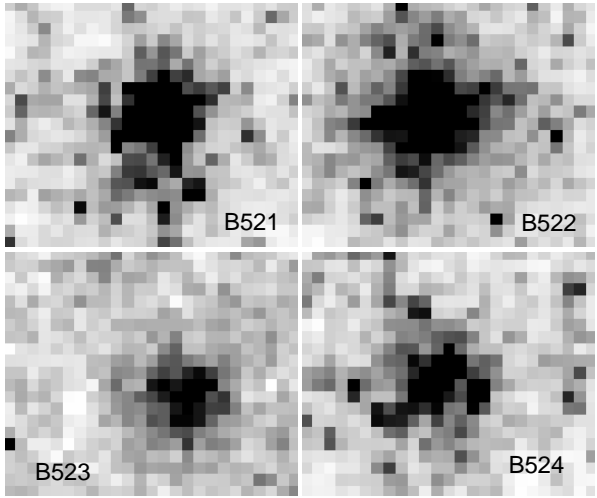}
    \caption{New globular cluster candidates found in HST images: B521,
B522, B523 and B524. The size of the post-stamp images shown here is
 $\sim 5\arcsec \times 5\arcsec $.}
           \label{imanew}
    \end{figure}

While inspecting this huge batch of HST images, we identified four new
candidates, listed in Tab.~\ref{hstnew} and christened B521, B522, B523,
and B524.
The images clearly reveal these objects as non-stellar: however they are faint
and not clearly resolved into stars, hence a spectrum is required to
ultimately assess their classification (see Fig.~\ref{imanew}).

During the observing runs at the Cassini Telescope (Sect.~2.2),
in addition to the remote CGCs listed in Tab.~\ref{oss}, we have obtained
radial velocities also for 7 candidates already listed in the RBC.
The results and classifications for these CGCs are listed in Tab.~\ref{rvold}.
All of these candidates turned out to be galaxies or foreground stars.

\begin{table}

\caption{Radial velocities of globular cluster candidates drawn from
the RBC \label{rvold}}

\begin{tabular}{l c r r r }
\hline \hline
Name  & $ V_r$  &$ \pm\epsilon_{Vr}$& Source&Classification\\
\hline
B052  & 30060 & 65   &  R3 LOI & galaxy             \\
B062  & 38180 & 53   &  R3 LOI & galaxy             \\
B503  & 29330 & 60   &27/09/05 & galaxy             \\
B285D & 14415 & 32   &  R2 LOI & galaxy             \\
B329D & 32141 & 51   &  R2 LOI & galaxy             \\
G289  & -53   &  7   &  R2 LOI & star?              \\
G295  & -74   & 12   &  R2 LOI & star?              \\
\hline
\end{tabular}
\end{table}

All the present observational material has been consistently implemented
to update the RBC, available on line in its latest V3.0 release.
Future minor updates of the catalog will be described and commented in the
RBC web page ({\tt http://www.bo.astro.it/M31/}) where the catalog
can be retrieved in form of ASCII files.

\end{appendix}

\onecolumn
\clearpage
\begin{small}

\begin{longtable}{l r r l l r l l r r }
 \caption{Globular clusters candidates from HST images \label{hst}} \\
\hline \hline
Name& S&V&Classification&c&Prop. ID&Camera&Filter&Dataset&Exptime \\    
\hline
\endfirsthead
\caption{continued.}\\
\hline\hline
Name& S&V&Classification&c&Prop. ID&Camera&Filter&Dataset&Exptime \\    
\hline
\endhead
\hline
\endfoot
B008  & C&  & globular cluster   &  1 &  10407  &  ASC/WFC  &F606W & j96q07010 &3250\\
B010  & C&  & globular cluster   &  1 &  10407  &  ASC/WFC  &F606W & j96q07010 &3250\\
B013  & C&  & globular cluster   &  1 &   9767  &  ACS/WFC  &F814W & j80420c9q & 774\\
B023  & C&C & globular cluster   &  1 &   9719  &  ACS/HRC  &F606W & j8pz02010 &2020\\
B037  & C&C & globular cluster   &  1 &  10260  &  ACS/WFC1 &F606W & j8z003010 &2370\\
B041  & C&  & globular cluster   &  1 &  10260  &  ACS/WFC1 &F606W & j8z003010 &2370\\
B042  & C&  & globular cluster   &  1 &  10260  &  ACS/WFC1 &F606W & j8z006010 &2370\\
B049  & C&  & globular cluster   &  1 &  10407  &  ASC/WFC  &F606W & j96q06010 &3250\\
B056  & C&  & globular cluster?  &  1 &   9087  &  ACS/WFC1 &F435W & j6d509020 &2100\\
B057  & C&  & globular cluster   &  1 &  10407  &  ASC/WFC  &F606W & j96q06010 &3250\\
B061  & C&C & globular cluster   &  1 &  10260  &  ACS/WFC1 &F606W & j8z008010 &2370\\
B063  & C&C & globular cluster   &  1 &  10260  &  ACS/WFC1 &F606W & j8z008010 &2370\\
B068  & C&  & globular cluster?  &  1 &   9087  &  ACS/WFC1 &F435W & j6d509020 &2100\\
B082  & C&  & globular cluster   &  1 &  10260  &  ACS/WFC1 &F606W & j8z004010 &2370\\
B086  & C&  & globular cluster   &  1 &  10094  &  ACS/WFC  &F814W & j92t46qpq & 824\\
B088  & C&  & globular cluster   &  1 &  10260  &  ACS/WFC1 &F606W & j8z007010 &2370\\
B090  & C&  & globular cluster?  &  1 &  10260  &  ACS/WFC1 &F606W & j8z004010 &2370\\
B091  & C&C & globular cluster?  &  1 &  10273  &  ACS/WFC  &F814W & j92gb3dnq & 502\\
B093  & C&C & globular cluster   &  1 &  10273  &  ACS/WFC  &F814W & j92gb3dnq & 502\\
B094  & C&  & globular cluster   &  1 &  10273  &  ACS/WFC  &F814W & j92gb8vwq & 502\\
B102  & C&A & asterism       &  7 &  10260  &  ACS/WFC1 &F606W & j8z007010 &2370\\
B103  & C&C & globular cluster   &  1 &  10006  &  ACS/WFC  &F435W & j8vp04010 &2200\\
B104  & C&  & globular cluster   &  1 &  10006  &  ACS/WFC  &F435W & j8vp04010 &2200\\
B112  & C&C & globular cluster   &  1 &  10006  &  ACS/WFC  &F435W & j8vp04010 &2200\\
B117  & C&  & globular cluster?  &  1 &   9087  & WFPC2/PC1 &F336W & j6d50105r & 500\\
B119  & C&  & globular cluster   &  1 &  10006  &  ACS/WFC  &F435W & j8vp04010 &2200\\
B120  &  &  & star       &  6 &  10006  &  ACS/WFC  &F435W & j8vp04010 &2200\\
B124  & C&C & globular cluster   &  1 &  10006  &  ACS/WFC  &F435W & j8vp03010 &2200\\
B126  & C&C & globular cluster   &  1 &  10006  &  ACS/WFC  &F435W & j8vp02010 &2200\\
B127  & C&C & globular cluster   &  1 &  10006  &  ACS/WFC  &F435W & j8vp03010 &2200\\
B128  &  &  & globular cluster?  &  2 &  10006  &  ACS/WFC  &F435W & j8vp02010 &2200\\
B130  & C&  & globular cluster   &  1 &  10273  &  ACS/WFC  &F814W & j92gb4e7q & 502\\
B131  & C&C & globular cluster   &  1 &  10006  &  ACS/WFC  &F435W & j8vp03010 &2200\\
B132  &  &  & globular cluster?  &  2 &  10006  &  ACS/WFC  &F435W & j8vp03010 &2200\\
B134  & C&  & globular cluster   &  1 &  10118  &  ACS/WFC  &F660N & j8zs04040 &2028\\
B140  & C&G & globular cluster   &  1 &  10273  &  ACS/WFC  &F814W & j92gb1fnq & 502\\
B144  & C&  & globular cluster   &  1 &  10118  &  ACS/WFC  &F814W & j8zs05a6q & 507\\
B145  &  &  & globular cluster?  &  2 &  10006  &  ACS/WFC  &F435W & j8vp02010 &2200\\
B146  & C&  & globular cluster   &  1 &  10118  &  ACS/WFC  &F814W & j8zs05a6q & 507\\
B147  & S&  & globular cluster   &  1 &  10260  &  ACS/WFC1 &F606W & j8z005010 &2370\\
B148  & C&C & globular cluster   &  1 &   9087  &  ACS/WFC1 &F435W & j6d508010 &2100\\
B151  & C&  & globular cluster   &  1 &  10260  &  ACS/WFC1 &F606W & j8z005010 &2370\\
B152  & C&C & globular cluster   &  1 &   9087  &  ACS/WFC1 &F435W & j6d508020 &2100\\
B153  & C&  & globular cluster   &  1 &   9087  &  ACS/WFC1 &F435W & j6d508010 &2100\\
B154  & C&C & globular cluster   &  1 &   9087  &  ACS/WFC1 &F435W & j6d508010 &2100\\
B155  & C&  & globular cluster   &  1 &   9480  &  ACS/WFC  &F775W & j8hohleeq & 700\\
B156  & C&  & globular cluster   &  1 &   9480  &  ACS/WFC  &F775W & j8hohleeq & 700\\
B158  & C&  & globular cluster   &  1 &   9719  &  ACS/HRC  &F606W & j8pz02010 &2020\\
B159  & C&  & globular cluster   &  1 &  10006  &  ACS/WFC  &F435W & j8vp08010 &2200\\
B160  & C&  & globular cluster   &  1 &   9480  &  ACS/WFC  &F775W & j8hohleeq & 700\\
B162  & C&C & globular cluster   &  1 &  10006  &  ACS/WFC  &F435W & j8vp08010 &2200\\
B169  & C&  & globular cluster   &  1 &  10407  &  ASC/WFC  &F606W & j96q03010 &3396\\
B171  & C&  & globular cluster   &  1 &  10407  &  ASC/WFC  &F606W & j96q03010 &3396\\
B174  & C&  & globular cluster   &  1 &  10273  &  ACS/WFC  &F814W & j92gb8vwq & 502\\
B175  & C&  & star       &  6 &  10407  &  ASC/WFC  &F606W & j96q03010 &3396\\
B185  & C&  & globular cluster   &  1 &  10407  &  ASC/WFC  &F606W & j96q03010 &3396\\
B198  & C&  & globular cluster   &  1 &  10407  &  ASC/WFC  &F606W & j96q05010 &1840\\
B199  & C&  & globular cluster   &  1 &   9392  & WFPC2/PC1 &F606W & u8f10201m &2400\\
B201  & C&  & globular cluster   &  1 &  10273  &  ACS/WFC  &F814W & j92gb6d3q & 502\\
B202  &  &  & globular cluster   &  1 &   9392  & WFPC2/PC1 &F606W & u8f10201m &2400\\
B203  & C&  & globular cluster   &  1 &  10407  &  ASC/WFC  &F606W & j96q05010 &1840\\
B206  & C&  & globular cluster   &  1 &  10407  &  ASC/WFC  &F606W & j96q05010 &1840\\
B213  & C&  & globular cluster   &  1 &  10407  &  ASC/WFC  &F606W & j96q05010 &1840\\
B215  & C&  & globular cluster   &  1 &  10407  &  ASC/WFC  &F606W & j96q05010 &1840\\
B220  & C&  & globular cluster   &  1 &  10407  &  ASC/WFC  &F606W & j96q02010 &1860\\
B224  & C&  & globular cluster   &  1 &  10407  &  ASC/WFC  &F606W & j96q02010 &1860\\
B225  & C&C & globular cluster   &  1 &   9719  &  ACS/HRC  &F606W & j8pz02010 &2020\\
B231  & C&  & globular cluster   &  1 &  10407  &  ASC/WFC  &F606W & j96q04010 &3315\\
B234  & C&  & globular cluster   &  1 &  10407  &  ASC/WFC  &F606W & j96q04010 &3315\\
B253  & C&  & stars      &  6 &  10407  &  ASC/WFC  &F606W & j96q06010 &3250\\
B257  &  &  & globular cluster?  &  2 &   9087  &  ACS/WFC1 &F435W & j6d509020 &2100\\
B261  &  &S & stars      &  6 &  10006  &  ACS/WFC  &F435W & j8vp04010 &2200\\
B264  & S&C & globular cluster?  &  2 &  10006  &  ACS/WFC  &F435W & j8vp03010 &2200\\
B353  & S&  & blank      &  6 &  10006  &  ACS/WFC  &F435W & j8vp03010 &2200\\
B366  & C&  & globular cluster   &  1 &  10407  &  ASC/WFC  &F606W & j96q01010 &1850\\
B367  & C&  & globular cluster   &  1 &  10407  &  ASC/WFC  &F606W & j96q01010 &1850\\
B407  & C&C & globular cluster   &  1 &   9458  &  WFPC2    &F814W & u8db0701m &1100\\
B458  & C&  & GC = B031D     &  1 &  10407  &  ASC/WFC  &F606W & j96q06010 &3250\\
B465  &  &  & blank      &  2 &  10118  &  ACS/WFC  &F814W & j8zs05a6q & 507\\
B515  &  &C & globular cluster   &  1 &  10260  &  ACS/WFC1 &F606W & j8z007010 &2370\\
B004D &  &  & galaxy         &  4 &  10407  &  ASC/WFC  &F606W & j96q07010 &3250\\
B034D & C&  & star       &  6 &  10407  &  ASC/WFC  &F606W & j96q06010 &3250\\
B056D &  &C & globular cluster   &  1 &  10260  &  ACS/WFC1 &F606W & j8z007010 &2370\\
B058D &  &  & stars      &  6 &  10273  &  ACS/WFC  &F814W & j92gb3dnq & 502\\
B059D &  &  & star       &  6 &  10273  &  ACS/WFC  &F814W & j92gb3cvq & 502\\
B062D &  &  & galaxy?        &  2 &  10273  &  ACS/WFC  &F814W & j92gb3cvq & 502\\
B065D &  &  & star       &  6 &  10006  &  ACS/WFC  &F435W & j8vp05010 &2200\\
B074D &  &S & blank      &  6 &  10006  &  ACS/WFC  &F435W & j8vp03010 &2200\\
B075D &  &S & star       &  6 &  10006  &  ACS/WFC  &F435W & j8vp03010 &2200\\
B077D &  &  & stars      &  6 &   9392  &  ACS/WFC  &F606W & j8f101mtq &1000\\
B080D &  &S & star       &  6 &  10118  &  ACS/WFC  &F814W & j8zs05a6q & 507\\
B081D &  &  & stars      &  6 &   9480  &  ACS/WFC  &F775W & j8hoiiq8q & 624\\
B083D &  &  & star       &  6 &   9480  &  ACS/WFC  &F775W & j8hoiiq8q & 624\\
B085D &  &  & stars      &  6 &   9392  &  ACS/WFC  &F606W & j8f101mtq &1000\\
B086D &  &S & star       &  6 &   9392  &  ACS/WFC  &F606W & j8f101mtq &1000\\
B087D &  &  & globular cluster?  &  2 &  10273  &  ACS/WFC  &F814W & j92gb1fnq & 502\\
B088D &  &  & globular cluster?  &  2 &   9719  &  ACS/WFC  &F625W & u8pz01020 &1448\\
B090D & C&  & globular cluster   &  1 &  10118  &  ACS/WFC  &F814W & j8zs05a6q & 507\\
B092D &  &  & stars?         &  2 &  10006  &  ACS/WFC  &F435W & j8vp02010 &2200\\
B093D &  &S & star       &  6 &   9392  &  ACS/WFC  &F606W & j8f101mtq &1000\\
B094D &  &  & stars      &  6 &  10006  &  ACS/WFC  &F435W & j8vp02010 &2200\\
B112D &  &  & globular cluster?  &  2 &   9719  &  ACS/WFC  &F625W & u8pz03020 &1448\\
B255D & C&  & globular cluster   &  1 &  10407  &  ASC/WFC  &F606W & j96q01010 &1850\\
B316D &  &  & star       &  6 &   9458  &   WFPC2   &F814W & u8db0801m &1100\\
B318D &  &  & star       &  6 &   9458  &   WFPC2   &F814W & u8db0801m &1100\\
B328D &  &  & galaxy         &  4 &   9458  &   WFPC2   &F814W & u8db0901m &1100\\
G137  &  &H & HII region     &  5 &  10260  &  ACS/WFC1 &F606W & j8z008010 &2370\\
G204  &  &  & star       &  6 &   9087  &  ACS/WFC1 &F435W & j6d508020 &2100\\
V254  &  &  & HII region?    &  2 &  10273  &  ACS/WFC  &F814W & j92gb6d3q & 502\\
NB17  & C&  & globular cluster?  &  1 &  10006  &  ACS/WFC  &F435W & j8vp05010 &2200\\
NB18  &  &  & blank      &  2 &  10006  &  ACS/WFC  &F435W & j8vp05010 &2200\\
NB21  & C&  & globular cluster?  &  1 &  10006  &  ACS/WFC  &F435W & j8vp05010 &2200\\
NB23  &  &  & globular cluster?  &  2 &  10006  &  ACS/WFC  &F435W & j8vp04010 &2200\\
NB26  & S&  & stars      &  6 &  10006  &  ACS/WFC  &F435W & j8vp02010 &2200\\
NB28  &  &  & star       &  6 &  10006  &  ACS/WFC  &F435W & j8vp05010 &2200\\
NB30  &  &  & star       &  6 &  10006  &  ACS/WFC  &F435W & j8vp03010 &2200\\
NB32  & S&  & stars      &  6 &  10006  &  ACS/WFC  &F435W & j8vp04010 &2200\\
NB34  &  &  & stars?         &  2 &  10118  &  ACS/WFC  &F814W & j8zs05a6q & 507\\
NB35  &  &  & globular cluster?  &  2 &  10006  &  ACS/WFC  &F435W & j8vp04010 &2200\\
NB36  &  &  & star       &  6 &  10006  &  ACS/WFC  &F435W & j8vp04010 &2200\\
NB37  &  &S & stars      &  6 &  10006  &  ACS/WFC  &F435W & j8vp04010 &2200\\
NB39  &  &  & star?      &  2 &  10006  &  ACS/WFC  &F435W & j8vp03010 &2200\\
NB40  &  &  & star       &  6 &  10118  &  ACS/WFC  &F660N & j8zs04040 &2028\\
NB42  &  &  & star       &  6 &  10006  &  ACS/WFC  &F435W & j8vp03010 &2200\\
NB44  &  &S & stars      &  6 &  10006  &  ACS/WFC  &F435W & j8vp05010 &2200\\
NB45  &  &S & stars      &  6 &  10118  &  ACS/WFC  &F660N & j8zs04040 &2028\\
NB49  &  &S & stars      &  6 &  10006  &  ACS/WFC  &F435W & j8vp02010 &2200\\
NB50  &  &  & star       &  6 &  10006  &  ACS/WFC  &F435W & j8vp02010 &2200\\
NB52  &  &  & star       &  6 &  10006  &  ACS/WFC  &F435W & j8vp02010 &2200\\
NB55  &  &  & star       &  6 &  10006  &  ACS/WFC  &F435W & j8vp03010 &2200\\
NB58  &  &  & star       &  6 &  10006  &  ACS/WFC  &F435W & j8vp03010 &2200\\
NB60  &  &  & blank      &  2 &  10006  &  ACS/WFC  &F435W & j8vp04010 &2200\\
NB68  & S&  & star       &  6 &  10006  &  ACS/WFC  &F435W & j8vp04010 &2200\\
NB70  & S&  & star       &  6 &  10006  &  ACS/WFC  &F435W & j8vp04010 &2200\\
NB73  &  &  & stars      &  6 &  10006  &  ACS/WFC  &F435W & j8vp04010 &2200\\
NB74  & S&  & star       &  6 &  10006  &  ACS/WFC  &F435W & j8vp03010 &2200\\
NB75  &  &S & star       &  6 &  10006  &  ACS/WFC  &F435W & j8vp03010 &2200\\
NB79  &  &  & star       &  6 &  10094  &  ACS/WFC  &F814W & j92t46qpq & 824\\
NB89  & C&  & globular cluster?  &  1 &  10006  &  ACS/WFC  &F435W & j8vp03010 &2200\\
NB90  &  &  & star       &  6 &  10118  &  ASC/WFC  &F660N & j8zs04040 &2028\\
NB92  &  &S & star?      &  6 &  10006  &  ACS/WFC  &F435W & j8vp02010 &2200\\
NB94  & S&S & star       &  6 &  10006  &  ACS/WFC  &F435W & j8vp03010 &2200\\
NB95  & S&S & star       &  6 &  10006  &  ACS/WFC  &F435W & j8vp03010 &2200\\
NB97  &  &S & star       &  6 &  10118  &  ACS/WFC  &F660N & j8zs04040 &2028\\
NB98  &  &  & star       &  6 &  10006  &  ACS/WFC  &F435W & j8vp02010 &2200\\
NB99  &  &S & star       &  6 &  10006  &  ACS/WFC  &F435W & j8vp02010 &2200\\
NB100 &  &S & star       &  6 &  10006  &  ACS/WFC  &F435W & j8vp02010 &2200\\
NB101 &  &  & star       &  6 &  10006  &  ACS/WFC  &F435W & j8vp02010 &2200\\
NB102 &  &  & star       &  6 &  10006  &  ACS/WFC  &F435W & j8vp04010 &2200\\
NB103 &  &S & star       &  6 &  10118  &  ACS/WFC  &F814W & j8zs05a6q & 507\\
NB104 &  &S & star       &  6 &  10006  &  ACS/WFC  &F435W & j8vp02010 &2200\\
NB105 &  &  & star       &  6 &  10006  &  ACS/WFC  &F435W & j8vp04010 &2200\\
NB106 &  &S & star       &  6 &  10006  &  ACS/WFC  &F435W & j8vp02010 &2200\\
AU007 &  &  & star       &  6 &  10006  &  ACS/WFC  &F435W & j8vp03010 &2200\\
AU008 &  &  & globular cluster?  &  2 &  10006  &  ACS/WFC  &F435W & j8vp05010 &2200\\
AU010 &  &  & globular cluster   &  1 &  10118  &  ACS/WFC  &F814W & j8zs05a6q & 507\\
M039  &  &  & faint      &  2 &  10407  &  ASC/WFC  &F606W & j96q02010 &1860\\
M047  &  &  & faint      &  2 &  10407  &  ASC/WFC  &F606W & j96q04010 &3315\\
M050  &  &  & faint, not star    &  2 &  10407  &  ASC/WFC  &F606W & j96q04010 &3315\\
M055  &  &  & globular cluster?  &  2 &   9767  &  ACS/WFC  &F606W & j8o440e9q & 774\\
BH16  &  &  & stars?         &  2 &  10006  &  ACS/WFC  &F435W & j8vp05010 &2200\\
BH18  &  &  & globular cluster?  &  2 &  10006  &  ACS/WFC  &F435W & j8vp02010 &2200\\
BH19  &  &  & globular cluster?  &  2 &   9392  &  ACS/WFC  &F606W & j8f101mtq &1000\\
BH25  &  &  & globular cluster?  &  2 &   9480  &  ACS/WFC  &F775W & j8hohleeq & 700\\
\end{longtable}
\end{small}

\end{document}